\documentclass[11pt]{article}
\usepackage{amsmath,amssymb,color,graphics,epsfig,cite}

\usepackage{amsfonts}

\newcommand{\be}{\begin{equation}}
\newcommand{\ee}{\end{equation}}
\newcommand{\bea}{\setlength\arraycolsep{2pt} \begin{eqnarray}}
\newcommand{\eea}{\end{eqnarray}}
\newcommand{\nn}{\nonumber}

\def\ft#1#2{{\textstyle{\frac{\scriptstyle #1}{\scriptstyle #2} } }}
\def\fft#1#2{{\frac{#1}{#2}}}

\def\0{{\sst{(0)}}}
\def\1{{\sst{(1)}}}
\def\2{{\sst{(2)}}}
\def\3{{\sst{(3)}}}
\def\4{{\sst{(4)}}}
\def\5{{\sst{(5)}}}
\def\6{{\sst{(6)}}}
\def\7{{\sst{(7)}}}
\def\8{{\sst{(8)}}}
\def\sst#1{{\scriptscriptstyle #1}}
\def\oneone{\rlap 1\mkern4mu{\rm l}}

\begin{document}

\begin{center}
{\Large {\bf Degenerate Black Rings in $D=5$ Minimal Supergravity}}

\vspace{20pt}

{\large Shi-Fa Guo, H. L\"u and Yi Pang}

\vspace{10pt}

{\it  Center for Joint Quantum Studies and Department of Physics,\\
School of Science, Tianjin University, Tianjin 300350, China}

\vspace{40pt}

\underline{ABSTRACT}
\end{center}

We consider both gauged and ungauged minimal supergravities in five dimensions and analyse the charged rotating solutions with two equal angular momenta $J$. When the electric charge $Q\sim J^{2/3}$ with some specific coefficient, we find new extremal black objects emerge that are asymptotic to either Minkowski or global AdS spacetimes and can be best described as degenerate black rings.  Their near-horizon geometry is locally AdS$_3\times S^2$, where the periodic $U(1)$ fibre coordinate in $S^3$ untwists and collapses to be the degenerate part of the AdS$_3$ horizon. It turns out that there are two branches of extremal rotating black holes, starting as the extremal RN black holes of the same mass, but opposite charges. With the increasing of the angular momentum, they will join to become the same degenerate black ring, where the Gibbs free energies however are not continuous at the joining. For the same $Q(J)$ relation, we find that there is in addition a rotating soliton whose mass is smaller than that of the degenerate black ring.

\vfill{\footnotesize shifaguo97@gmail.com  \ \ \ mrhonglu@gmail.com \ \ \ pangyi1@tju.edu.cn}



\thispagestyle{empty}
\pagebreak

\section{Introduction}

Static and spherically symmetric solutions in General Relativity are relatively easy to construct and in many cases the equations can be reduced to some quadratures, if not exactly solvable. For example, the Reissner-Nordestr\o m (RN) black hole can be straightforwardly generalized to higher dimensions, with or without a cosmological constant. The situations for rotating geometries are much more complicated. While the Kerr metric \cite{Kerr:1963ud} and its neutral Plebanski extension \cite{Plebanski:1975xfb} can be generalized to higher dimensions \cite{Myers:1986un,Hawking:1998kw,Gibbons:2004uw,Chen:2006xh}, the generalization of the (charged) Kerr-Newman black hole \cite{Newman:1965my} to Einstein-Maxwell gravity in higher dimensions, on the other hand, has not been successful.

   The situation improves with supergravities, the low energy effective theories of strings.
These theories typically have enhanced global symmetry, which makes it easier to construct new solutions. Charged rotating solutions in supergravities have been obtained through solution generating techniques, utilising the enhanced global symmetries in lower dimensions, see e.g.~\cite{Sen:1992ua,Cvetic:1996xz,Cvetic:1996dt}. Even though these global symmetries are broken in gauged supergravities where scalar potentials are introduced, many charged rotating AdS black holes were nevertheless constructed, e.g.~\cite{Chong:2005hr,Wu:2011gq}. These solutions involve mass, charge and angular momenta and are typically very complicated. For suitable such parameters, they describe black holes. However, the spacetime geometries for general parameters have not been thoroughly investigated.

Rotating geometry is particularly important in higher dimensions where black objects with new topologies other than spheres can arise. An important breakthrough in black hole research is the discovery of five-dimensional black rings whose horizon topology is not $S^3$, but $S^1\times S^2$ \cite{Emparan:2001wn,Pomeransky:2006bd}, and hence they are necessarily rotating. Supersymmetric black rings in five dimensional supergravities have since been constructed \cite{Elvang:2004rt}.  All known black rings are asymptotically flat. There are no known examples of black rings that are asymptotic to the global anti-de Sitter (AdS) spacetimes. In particular, a no-go theorem was established that a supersymmetric AdS black ring is not possible \cite{Grover:2013hja}. The no-go theorem was further extended to the nonexistence of general extremal de Sitter black rings \cite{Khuri:2017zqg}.

In this paper, we focus on minimal supergravity in five dimensions, both gauged and ungauged, and study the global structures of the charged rotating solutions that were constructed in \cite{clpd5sol1}.  The solution contains mass $M$, two equal angular momenta $J$ and electric charge $Q$.  For suitable choices of these parameters, we find new spacetimes that can be best described as degenerate black rings (DBRs). These solutions are asymptotic to Minkowski or global AdS spacetimes. They are extremal black objects with the near-horizon geometry of locally AdS$_3\times S^2$.  Specifically, the level surfaces of the spatial section outside the horizon are $S^3$, written as a $U(1)$ bundle over $S^2$. The $U(1)$ fibre untwists and collapses on the horizon and becomes the degenerate part of the AdS$_3$ factor. It appears that we have lost one dimension on the horizon with the radius of the $U(1)$ circle shrinking to zero. Unlike the usual black rings, our DBRs emerge in both gauged and ungauged supergravities.

The paper is organized as follows.  In section \ref{sec:local}, we review the general charged rotating solutions in minimal gauged supergravity in five dimensions. We list possibilities of global structures that can emerge from these solutions.  In section \ref{sec:dring}, we give the explicit solutions that describe DBRs. We analyse the near-horizon geometries of AdS$_3\times S^2$.  In section \ref{sec:phase}, we observe that the original local solution has two branches of extremal black holes. They start at one end as the RN black holes of equal mass, but opposite charges, and join at the other end as the DBR.  We show that there is a global discontinuity at the joining where the Gibbs free energies are not continuous.  In section \ref{sec:soliton}, we notice that the soliton solutions and DBRs have the same charge/angular momentum relation. We can thus study the DBRs from the perspective of the solitons.  We conclude the paper in section \ref{sec:conclusion}.  In appendix \ref{app}, we present the global analysis for the black holes and time machines that emerge from the local solutions.

\section{The local solution}
\label{sec:local}

The bosonic sector of minimal (${\cal N}=2$) gauged supergravity in five dimensions consists of the metric $g_{\mu\nu}$ and the graviphoton $A_\mu$. The Lagrangian 5-form is
\be
{\cal L} = (R-2\Lambda) {*\oneone} -\fft12 {*F\wedge F} + \fft{1}{3\sqrt3} F\wedge F\wedge A\,,
\ee
where $F=dA$.  The theory has a negative cosmological constant $\Lambda=-6g^2$, where $g$ is the gauge coupling constant of the gravitino $\psi_\mu$, which we set zero in this paper. It follows that $\ell=1/g$ is the radius of the AdS vacuum. Setting $g=0$ gives rise to ungauged minimal supergravity with Minkowski vacuum. A general local solution describing charged rotating AdS black hole with two equal angular momenta was constructed in \cite{clpd5sol1}, given by
\be
ds^2 = -\fft{f}{W} dt^2 + \fft{dr^2}{f} + \ft14 r^2 W (\sigma_3 + \omega dt)^2 +
\ft14 r^2 d\Omega_2^2\,,\qquad A = -\frac{\sqrt{3} a q}{2 r^2} \Big(\sigma_3 - \fft{2}{a} dt\Big)\,,\label{localsol}
\ee
where
\bea
W &=& 1+\frac{2 a^2 (\mu +q)}{r^4}-\frac{a^2 q^2}{r^6}\,,\qquad
\omega = \frac{2 a \left(q^2-q r^2-2 \mu  r^2\right)}{r^6W}\,,\nn\\
f &=& W\left( 1- (1-a^2 g^2)\fft{r^2}{a^2}\right)+\frac{\left(r^4+a^2 q\right)^2}{a^2 r^6}\,.
\eea
Note that the spatial level surfaces are 3-sphere that is written as a $U(1)$ bundle $\sigma_3$ over $S^2$, with the round unit $S^3$ metric:
\be
d\Omega_3^2 = \ft14 \sigma_3^2 + \ft14 d\Omega_2^2\,,\qquad
\sigma_3 = d\psi + \cos\theta d\phi\,,\qquad d\Omega_2^2 = d\theta^2 + \sin^2\theta d\phi^2\,.
\ee
The $U(1)$ coordinate $\psi$ has a period $\Delta \psi=4\pi$.  We shall also consider in some part of the paper the more general lens spaces $S^3/{\mathbb Z}_k$, in which case, we have
\be
\Delta \psi = \fft{4\pi}{k}\,,\qquad k=1,2,3,\cdots\,.\label{psiperiod}
\ee
Note that the original solution in \cite{clpd5sol1} has an additional parameter $\beta$, which was shown to be redundant \cite{Madden:2004ym}. We rename and set the parameters $(M,J,Q,\beta)$ of \cite{clpd5sol1} to be $(\mu, a, q,0)$ here.

The Riemann tensor squared is given by
\bea
&&\ft14 r^{16}\, R^{\mu\nu\rho\sigma}R_{\mu\nu\rho\sigma}=\nn\\
&&34 a^4 q^4-12 r^6 \left(8 a^2 (2 \mu +q)^2+15 \mu  q^2\right)-2 a^2 q^2 r^2 \left(192 a^2 (\mu +q)+113 q^2\right)\nn\\
&&+r^4 \left(384 a^4 (\mu +q)^2+96 a^2 q^2 (7 \mu +4 q)+127 q^4\right)+72 \mu ^2 r^8 + 2 g^2 r^2 \Big(113 a^4 q^4\nn\\
&&-2 a^2 r^6 \left(36 \mu ^2+13 q^2+36 \mu  q\right)+6 a^2 r^4 \left(32 a^2 (\mu +q)^2+15 q^2 (2 \mu +q)\right)\nn\\
&&-a^2 q^2 r^2 \left(336 a^2 (\mu +q)+127 q^2\right)+q^2 r^8\Big) +
g^4 r^4 \Big(127 a^4 q^4-180 a^4 q^2 r^2 (\mu +q)\nn\\
&&+72 a^4 r^4 (\mu +q)^2-2 a^2 q^2 r^6+10 r^{12}\Big)\,.
\eea
The spacetime thus has a curvature power-law singularity at $r=0$.  The metric approaches five-dimensional  Minkowski ($g^2=0$), global AdS ($g^2>0$) or static de Sitter ($g^2<0$) spacetimes asymptotically as $r\rightarrow \infty$:
\be
ds_5^2 = -(g^2 r^2 + 1) dt^2 + \fft{dr^2}{g^2 r^2 + 1} +  r^2 d\Omega_3^2\,.
\ee
In some complicated solutions such as \eqref{localsol}, the same local solution can describe different manifolds in disconnected regions. For example, over rotating black holes in one choice of radial coordinate can be mapped to under rotating ones in a different set of coordinates \cite{Chen:2006ea}.  For simplicity, we shall consider the spacetime in \eqref{localsol} where $r$ is only positive and the 3-sphere coordinates are fixed.

The solution \eqref{localsol} has three integration constants $(\mu,a,q)$, parameterizing the mass, angular momentum and electric charge:
\be
M=\ft14\pi \Big(3 \mu+ g^2 a^2 (\mu +q)\Big)\,,\qquad
J=\ft{1}{4} \pi  a (2 \mu +q)\,,\qquad Q=\ft{1}{4} \sqrt{3} \pi  q\,.\label{MJQ}
\ee
The mass is calculated following the steps of \cite{Chen:2005zj}, based on the AMD formalism \cite{am,ad}.  The angular momentum and electric charge are defined by
\bea
J &=& \fft{1}{16\pi} \int_{S^3} {*dK}\,,\quad \hbox{with}\quad K=2\fft{\partial}{\partial \psi}\,,\nn\\
Q &=& \fft{1}{16\pi} \int_{S^3} \Big({*F} - \fft{1}{\sqrt3} A\wedge F\Big)\,.
\eea
Note that these quantities are presented assuming the level surfaces are $S^3$. If we consider the lens spaces $S^3/{\mathbb Z}_k$ instead, the extensive quantities \eqref{MJQ} should all be divided by the natural number $k$:
\be
M_k=\fft{M}{k}\,,\qquad J_k = \fft{J}{k}\,,\qquad Q_k= \fft{Q}{k}\,.\label{mjqk}
\ee

Depending on the parameters, the local solution \eqref{localsol} can describe a variety of objects with very different topology and spacetime structures.  If the function $f$ has no real (positive) root, the curvature singularity at $r=0$ is naked.  For appropriate parameters, there are real roots for $f$ and we assume that the largest one is $r_+$, which must be positive since only even powers of $r$ appear in $f$.

The metric is degenerate at $r=r_+$, and the property of the corresponding collapsed curves depends also on the behavior of the function $W$. While the function $f$ could be absent from a real root, there must exist a real root for $W$ when the angular momentum is nonvanishing, since $W(\infty)=1$ and $W(0)=-\infty$.  We refer to the largest root of $W$ as $r_L$.  It is important to note that $r=r_L$ is not a coordinate singularity, but the velocity of light surface (VLS),  a timelike hypersurface boundary inside which the periodic $U(1)$ coordinate $\psi$ becomes timelike, and hence the  closed timelike curves (CTCs) develop \cite{Cvetic:2005zi}.  We consider the case where $r_+$ exists, three possibilities were proposed \cite{Cvetic:2005zi}
\bea
r_L<r_+:&& \hbox{black holes;}\nn\\
r_L>r_+:&& \hbox{time machines;}\nn\\
r_L=r_+:&& \hbox{solitons.}\label{first3}
\eea
In the appendix, we give the global analysis for the first two cases, and present some discussions on the solitons in section \ref{sec:soliton}. In this paper, we consider a fourth possibility:
\be
f(r_+)=f'(r_+)=W(r_+)=0\,.\label{dbrcond}
\ee
In other words, $r_L=r_+$ is a double root for $f$, but a single root for $W$. The solution is neither a black hole, nor a soliton but something that can be best described as an extremal degenerate black ring (DBR).  We shall discuss this in the next section.

\section{Degenerate black rings}
\label{sec:dring}

In this section, we present our main result, the DBR solutions that arise from the condition
\eqref{dbrcond}. Since the solution is particularly simple when $\Lambda=-6g^2=0$, we shall analyse this case first and then give the more general $g\ne 0$ later.

\subsection{$\Lambda = 0$}

When $g=0$, the constraint \eqref{dbrcond} implies $\mu=-q=a^2$.  The mass, charge and angular momentum reduce to
\be
M=-\sqrt3 Q = \frac{3 \sqrt[3]{\pi }}{2^{2/3}} J^{\fft23}\,,\qquad
J=\ft18 \pi a^3\,.\label{mjqring1}
\ee
The solution simplifies significantly
\bea
f&=& \left(1-\frac{a^2}{r^2}\right)^2\,,\qquad W=1-\frac{a^6}{r^6}\,,\qquad
\omega=-\frac{2 a^3}{a^4+a^2 r^2+r^4}\,,\nn\\
A &=& \fft{\sqrt3 a^3}{2r^2} \tilde \sigma_3\,,\qquad \tilde \sigma_3 = \sigma_3 - \fft{2}{a} dt\,.
\eea
In terms of $\tilde \sigma_3$, the metric because even simpler:
\be
ds^2 = \fft{r^4 dr^2}{(r^2 - a^2)^2} - \Big(1-\fft{r^2}{a^2}\Big) dt^2 +
\fft{r^4-a^4}{a^2 r^2} dt \tilde \sigma_3 + \fft{r^6 - a^6}{4a^2 r^4}\tilde \sigma_3^2 +
\ft14 r^2 d\Omega_2^2\,.
\ee
The metric is asymptotic to the Minkowski spacetime. The horizon is located at $r=a$ and it is instructive to define a new radial coordinate $z$, such that $r=a(1+z^2)$.  As $z$ approaches zero, the solution, up to the next to the leading order of $z$, becomes
\bea
ds^2 &\sim & \fft{a^2 dz^2}{z^2(1-3z^2)} + z^2 \left(- \ft13 (2 + z^2) dt^2 +
\ft34 a^2 (2-3z^2) \Big(\tilde \sigma_3 + \frac{4 (1+z ^2)}{3 a}dt\Big)^2\right)\nn\\
&& + \ft14 a^2 (1 + 2z^2) d\Omega_2^2\,,\nn\\
A &\sim & \ft{1}{2} \sqrt{3} a \left(1-2 z ^2\right) \tilde \sigma_3\,.\label{metricrho}
\eea
On the horizon, the solution carries a magnetic dipole moment
\be
{\cal D} = \fft{1}{8} \int F = \ft14 \sqrt3\,\pi a\,.
\ee

The near-horizon geometry is locally AdS$_3\times S^2$, since $\tilde \sigma_3$ is a periodic 1-form. The near-horizon geometry can be magnified and extracted as a solution on its own, under the decoupling limit
\be
r=a (1 + \lambda z^2)\,,\qquad
t=\fft{a\tau}{\sqrt{\lambda}}\,,\qquad
\tilde \psi = \fft{1}{\sqrt{\lambda}} \Big(\hat \psi + \ft23 \tau\Big)\,,\qquad
\lambda\rightarrow 0\,.\label{decoupling}
\ee
In this limit, the solution becomes
\bea
ds^2 &=& a^2 \Big(\fft{dz^2}{z^2} + \ft16 z^2 (-4 d\tau^2 + 9 d\hat \psi^2) +
\ft14 (d\theta^2 + \sin^2\theta d\phi^2)\Big)\,,\nn\\
A &=& \ft12\sqrt3 a \cos\theta d\phi\,.\label{brhorizon}
\eea
The AdS$_3$ factor of the metric is written in the planar coordinates. The solution \eqref{brhorizon} resembles the near-horizon geometry of the magnetic string, but our solution should not be described as a string. We need to distinguish the near horizon geometry \eqref{metricrho} and the decoupling limit \eqref{brhorizon}.  In the latter case, the $U(1)$ fibre is infinitely magnified and its global $U(1)$ property is lost.  Nevertheless it does show that the $U(1)$ fibre over the $S^2$ becomes untwisted in the AdS throat. In addition, five-dimensional strings would have the two-dimensional world volume and three-dimensional transverse space and the asymptotic spacetime is either Mink$_4\times S^1$ or Mink$_4\times \mathbb R$, depending whether the string is circular or a real line. In our solution, it is clear from the asymptotic Minkowski structure that the transverse space is four dimensional, as in the case of an electrically charged black hole. The solution should be best described as a degenerate black ring.  In a usual black ring, the horizon is $S^1\times S^2$, and in the near horizon geometry the $S^1$ has finite radius and it bundles over $\mathbb{R}^2$ or AdS$_2$ in the extremal case to form AdS$_3$, but asymptotically, it bundles over $S^2$ to form $S^3$ instead. In our solution, the radius of the $S^1$, i.e. the ring size, shrinks to zero on the horizon, and it appears as if we have lost one spatial dimension viewed from the asymptotic region.

It is worth commenting that the metric (\ref{metricrho}) is invariant under
\be
z\leftrightarrow -z\,.
\ee
If the AdS$_3$ horizon is geodesically complete, the solution would be regular with no singularity, since the inside of the horizon is isomorphic to the outside.  This would be the
case of the magnetic string where $\psi$ describes a real line \cite{Gibbons:1994vm}.  However, in our case, $\psi$ is periodic and hence the translational symmetry becomes singular on the horizon of the AdS$_3$. Thus our DBR solutions are not regular and the curvature singularity at $r=0$ can be reached geodesically. However, we can introduce
an orbifold singularity on the AdS$_3$ horizon so that the coordinate $z^2\ge 0$. This is a small price to pay to avoid the power-law singularity completely.

It is important to note that the charge/angular momentum relation is completely fixed:
\be
Q = -\frac{\sqrt{3} \sqrt[3]{\pi }}{2^{2/3}} J^{2/3}\,.\label{Q(J)}
\ee
As we shall see in section \ref{sec:soliton} that for the same relation, there exists further a smooth soliton that is asymptotic also to the Minkowski spacetime.  It is sensible to compare their masses for given $Q(J)$ and we find
\be
\fft{M_{\rm soliton}}{M_{\rm ring}} = \cos\left(\fft{\pi}{9}\right)<1\,.
\ee
In other words, the soliton is more energetically favoured than the DBR.

\subsection{$\Lambda \ne 0$}

There is no known exact solution of black rings that are asymptotic to global AdS. Our AdS DBRs however arise naturally. The constraint \eqref{dbrcond} with nonvanishing $g$ yields
\be
\mu=\frac{1}{2} r_+^2 \left(2+3 g^2 r_+^2+g^4 r_+^4\right)\,,\qquad
q=-r_+^2 \left(g^2 r_+^2+1\right)\,,\qquad a=\frac{r_+}{\sqrt{1+g^2 r_+^2}}\,.
\ee
The solution now reduces to
\bea
f&=& \Big(1 - \fft{r_+^2}{r^2}\Big)^2\Big(1+ g^2 \left(r^2+2 r_+^2\right)\Big)\,,\qquad
W=\Big(1 - \fft{r_+^2}{r^2}\Big)\Big(1 +
\fft{r_+^2}{r^2} + \frac{r_+^4 \left(g^2 r_+^2+1\right)}{r^4}\Big)\,,\nn\\
\omega &=& -\frac{2 r_+^3 \left(g^2 r_+^2+1\right)^{3/2}}{g^2 r_+^6+r^4+r^2 r_+^2+r_+^4}\,,\qquad A=\frac{r_+^3 \sqrt{3 g^2 r_+^2+3}}{2 r^2} \tilde \sigma_3\,.
\eea
Here $f$ has a double root $r_+$ and $W$ has the same single root.  The solution is asymptotic to the global AdS$_5$. The mass, charge and angular momentum now depend on $r_+$, given by
\bea
M &=&\frac{1}{8} \pi  r_+^2 \left(6+9 g^2 r_+^2+4 g^4 r_+^4\right)\,,\qquad
Q = -\frac{1}{4} \sqrt{3} \pi  r_+^2 \left(1+g^2 r_+^2\right)\,,\nn\\
J&=&\frac{1}{4} \pi  r_+^3 \left(1+g^2 r_+^2\right)^{3/2}.
\eea
The near horizon geometry is again locally AdS$_3\times S^2$, which can be blown up to be a solution on its own under the same decoupling limit $\lambda\rightarrow 0$, with
\bea
r &=& r_+(1 + \lambda z^2)\,,\qquad
t=\fft{r_+ \sqrt{1 + \fft13 g^2 r_+^2}}{1 + 3 g^2 r_+^2} \fft{\tau}{\sqrt{\lambda}}\,,\nn\\
\psi&=&\fft{1}{ \sqrt{\lambda(1 + \fft13 g^2 r_+^2)} (1 + 3 g^2 r_+^2)}\Big(\hat\psi +
\fft{2(1 + g^2 r_+^2)^{\fft32}}{3\sqrt{1 + 3 g^2 r_+^2}} \tau\Big).
\eea
The solution becomes
\bea
ds^2 &=& \fft{r_+^2}{1 + 3g^2 r_+^2} \Big(\fft{dz^2}{z^2} + \ft16 z^2 (-4 d\tau^2 + 9 d\hat \psi^2)\Big) +
\ft14 r_+^2 (d\theta^2 + \sin^2\theta d\phi^2)\,,\nn\\
A &=& \ft12r_+ \sqrt{3(1 + g^2 r_+^2)}\, \cos\theta d\phi\,.\label{brhorizon2}
\eea
Note that if the $\hat \psi$ is a real line, then the corresponding magnetic string solution would not be asymptotic to AdS, but locally AdS.  Here the coordinate has an origin of the $U(1)$ fibre of the $S^2$, and the DBR is indeed asymptotic to the global AdS. As in the earlier $g=0$ case, the AdS DBR solution can avoid the curvature power-low singularity at $r=0$ by introducing an orbifold singularity on the horizon.

The charge/angular momentum relation $Q(J)$ is independent of $g$, give by \eqref{Q(J)}. The mass on the other hand becomes more complicated, given by
\bea
M &=& \frac{\left(\sqrt{8 \sqrt[3]{2} g^2 J^{2/3}+\pi ^{2/3}}-\sqrt[3]{\pi }\right) \left(16 \sqrt[3]{2} g^2 J^{2/3}+5 \sqrt[3]{\pi } \sqrt{8 \sqrt[3]{2} g^2 J^{2/3}+\pi ^{2/3}}+7 \pi ^{2/3}\right)}{32 g^2}\nn\\
&=&
\left\{
  \begin{array}{ll}
    \frac{3 \sqrt[3]{\pi } J^{2/3}}{2^{2/3}}+\frac{3 g^2 J^{4/3}}{\sqrt[3]{2 \pi }}-\frac{4 g^4 J^2}{\pi }+O\left(J^{8/3}\right), &\qquad J\rightarrow 0\,; \\
    2 g J+\frac{3 \sqrt[3]{\pi } J^{2/3}}{2\ 2^{2/3}}+\frac{3 \pi ^{2/3} \sqrt[3]{J}}{8 \sqrt[3]{2} g}-\frac{\pi }{16 g^2}+O\left(J^{-\fft13}\right), &\qquad J\rightarrow \infty\,.
  \end{array}
\right.\label{mjadsring}
\eea
For the same charge/angular momentum relation \eqref{Q(J)}, there also exist AdS solitons and we shall come back to the discussion of the mass in section \ref{sec:soliton}.

\section{DBRs as limiting extremal rotating black holes}
\label{sec:phase}

In the previous section, we gave the DBR solutions that arise from the condition \eqref{dbrcond}.  We can impose this condition in two stages. The first is to set $f(r_+)=f'(r_+)=0$, which gives rise to the extremal rotating black holes, followed by requiring $W(r_+)=0$.  Thus the DBRs can be viewed as the limiting case of black holes.
It turns out that there are two branches of extremal rotating black holes and the DBR sits where they meet. The situations are sufficiently different whether $\Lambda=-6g^2$ vanishes or not and we discuss them separately.

\subsection{$\Lambda =0$}

The general non-extremal black hole solutions are discussed in the appendix, where the black hole thermodynamical variables are all given and the first law is derived.  We find that the extremal condition $f'(r_+) = f(r_+) =0$ can be satisfied by two sets of parameters.  For $g=0$, they are
\bea
\hbox{Case 1:}&& \quad \mu=r_+^2\,,\qquad q=-r_+^2\,,\nn\\
\hbox{Case 2:}&&\quad \mu=r_+^2\,,\qquad q=-2a^2 + r_+^2\,.
\eea
The first case is the BMPV black hole with \cite{Breckenridge:1996is}
\be
f=\Big(1 - \fft{r_+^2}{r^2}\Big)^2\,,\qquad
W=1 - \fft{a^2 r_+^4}{r^6}\,,\qquad \omega = -\frac{2 a r_+^2 \left(r^2-r_+^2\right)}{r^6-a^2 r_+^4}\,.
\ee
The solution is supersymmetric, with
\be
M=-\sqrt3 Q=\frac{3 \pi r_+^2}{4}\,,\qquad J=\frac{1}{4} \pi  a r_+^2\,.
\ee
In other words, the angular momentum does not contribute to the mass. Since the black hole has zero temperature, the Helmholtz free energy is simply the mass. We can further define three types of Gibbs energy
\be
G_1= M - 2\Omega_+ J\,,\qquad G_2 = M - 2\Omega_+ J - \Phi Q\,,\qquad
G_3 = M - \Phi Q\,.\label{gibbsdef}
\ee
For the BMPV black hole, we find
\be
G_1 = \ft34\pi r_+^2\,,\qquad G_2=G_3=0\,.
\ee
Note that these free energies are completely independent of the parameter $a$, and therefore $J$.

We now examine the second case, for which the solution is
\bea
f&=&\Big(1 - \fft{r_+^2}{r^2}\Big)^2\,,\qquad
W=1-\frac{4 a^2 \left(a^2-r_+^2\right)}{r^4}-\frac{a^2\left(r_+^2-2 a^2\right)^2}{r^6}\,,\nn\\
\omega &=& \frac{r^2 \left(4 a^3-6 a r_+^2\right)+2 a \left(r_+^2-2 a^2\right)^2}{r^6-4 a^2 r^2 \left(a^2-r_+^2\right) -\left(a r_+^2-2 a^3\right)^2}\,.
\eea
The mass, angular momentum and charge are
\be
M=\ft34 \pi  r_+^2\,,\qquad J=\ft{1}{4} \pi  a \left(3 r_+^2-2 a^2\right)\,,\qquad Q=\ft{1}{4} \sqrt{3} \pi  \left(r_+^2-2 a^2\right)\,.
\ee
The solution is non-supersymmetric and the conserved quantities satisfy
\be
8M^3 -18 M Q^2 - 6\sqrt3 Q^3 - 27\pi J^2=0\,.
\ee
The Gibbs free energies for this solution are
\be
G_1=\frac{\pi  \left(8 a^4-6a^2 r_+^2 +3 r_+^4\right)}{4 \left(r_+^2 + 2 a^2\right)}\,,\qquad
G_2=\frac{\pi  a^2 \left(3 r_+^2-2 a^2\right)}{2(r_+^2 + 2 a^2)}\,,\qquad G_3=\frac{3 \pi  a^2 \left(3 r_+^2-2 a^2\right)}{2 \left(r_+^2 + 2 a^2\right)}\,.
\ee

At this stage, it should be stated that both Case 1 and Case 2 solutions describe rotating black holes only when $r_+>a$, for which $W(r_+)>0$.  When $r_+<a$, we have negative $W(r_+)$ and the solution describes a time machine, which was referred to as a repulson in \cite{Gibbons:1999uv}.  Interestingly, when $r_+=\sqrt{3/2} a$, the angular momentum vanishes, we have a rotating time machine repulson with no angular momentum.

When $r_+=a$, the two different solutions become the same DBR studied in the previous section.  The DBR sits in the same crossing between the black holes and time machines. While the Gibbs free energies are continuous when the black hole transiting to the time machine in each case, they are not continuous when transiting from Case 1 to Case 2 via the DBR. When $r_+=a$, the Case 2 solution has $G_{1,2,3} = \left\{\frac{5 \pi  a^2}{12},\frac{\pi  a^2}{6},\frac{\pi  a^2}{2}\right\}$, which are all bigger than the counterparts of the black ring limit of the Case 1 solution. This global discontinuity of black hole thermodynamics is rather peculiar and deserves further investigation.

   We can also approach the black hole thermodynamics by the on-shell Euclidean action using
the quantum statistic relation. In particular, the Gibbs free energies $G_1$ and $G_2$ defined in \eqref{gibbsdef} can be derived from the Euclidean action with appropriate asymptotic boundary terms of the Maxwell field. This leads to an apparent paradox, since the solutions are well defined on and outside of the horizon and therefore there is no obvious reason for the  discontinuity in the Euclidean action. The key lies in the fact that the $FFA$ term in the action is not gauge invariant, even though its contribution to the equations of motion is. The correct Euclidean action associated with the Gibbs free energies require that the quantity $\ell^\mu A_\mu$ must vanish on the horizon rather than at infinity, where $\ell^\mu$ is the null Killing vector on the horizon.  This ensures that $A^2$ is well defined on and outside of the horizon. With this requirement, one can show that at the DBR limit, the temporal component of the gauge field in the two branches of extremal black holes differ from each other by a constant shift, thus leading to different values for the Gibbs free energies.

\subsection{$\Lambda\ne 0$}

The situation is analogous when the cosmological constant is turned on with $g\ne 0$, but the details become more complicated.  The condition $f(r_+)=0=f'(r_+)$ for extremal black holes again yields two sets of solutions
\be
q_\pm=-\fft{1}{(1-a^2 g^2)^2}\left(a^2 \pm  (r_+^2 \left(1-a^2 g^2\right)-a^2)
\sqrt{1+ 2 g^2 r_+^2 \left(1-a^2 g^2\right)}\right)\,,\label{qpm}
\ee
with
\be
\mu = \fft{2 a^2 g^2 q+r_+^2 \left(3 g^2 r_+^2+2\right)}{2(1-a^2 g^2)}\,.
\ee
The plus and minus signs in the parenthesis in the above $q$ expression correspond respectively to the Case 1 and Case 2 solutions when $g=0$ and we shall use the same names here. As in the case of $g=0$, the metric function $f$ is the same for both solutions
\be
f=\Big(1- \fft{r_+^2}{r^2}\Big)^2 \Big(1 + g^2 (r^2 + 2 r_+^2)\Big)\,.
\ee
The functions $W$ and $\omega$ depend linearly on $q$ and give two different branches of solutions
\bea
W &=& 1 + \frac{2 a^2 q \left(a^2 + (1-a^2 g^2) r^2\right)}{r^6 \left(1-a^2 g^2\right)^2}+\frac{a^2 r_+^2}{r^6 \left(1-a^2 g^2\right)^2} \Bigg(r^2 \left(1-a^2 g^2\right) \left(3 g^2 r_+^2+2\right)\nn\\
&&+2 a^2 \left(g^2 r_+^2+1\right)^2-r_+^2 \left(2 g^2 r_+^2+1\right)\Bigg)\,,\nn\\
\omega &=& \fft{2a}{r^6 W} \Bigg(\frac{((a^4 g^4-1) r^2-2 a^2)q}{\left(1-a^2 g^2\right)^2}-
\frac{r^2 r_+^2 \left(3 g^2 r_+^2+2\right)}{1-a^2 g^2}\nn\\
&&+\frac{r_+^4+2 g^2 r_+^6 -2 a^2r_+^2 \left(g^2 r_+^2+1\right)^2}{\left(1-a^2 g^2\right)^2}\Bigg)\,.
\eea
The two generally different black holes of (\ref{qpm}) become the same DBR discussed in the previous section when $r_+$ is
\be
r_+ = \frac{a}{\sqrt{1-a^2 g^2}}\,.\label{rpag}
\ee
However, the Gibbs free energies of the two rotating black holes are not the same when we take
this limit. They are given by
\bea
G_1^\pm &=& \frac{\pi  a^2 \left(2 a^4 g^4-7 a^2 g^2+14
\pm 4 \sqrt{2 a^2 g^2+1}\right)}{8 \left(1-a^2 g^2\right)^2 \left(3-2 a^2 g^2\right)}\,,\nn\\
G_2^\pm &=& \frac{\pi  a^2 \left(2 a^4 g^4-7 a^2 g^2+2 \mp 2 \sqrt{2 a^2 g^2+1}\right)}{8 \left(1-a^2 g^2\right)^2 \left(3-2 a^2 g^2\right)}\,,\nn\\
G_3^\pm &=& \frac{\pi  a^2 \left(6 - a^2 g^2 (3-a^2 g^2)(3-2a^2 g^2) \mp 6 \left(1-a^2 g^2\right) \sqrt{2 a^2 g^2+1} \right)}{8 \left(1-a^2 g^2\right)^3 \left(3-2 a^2 g^2\right)}\,.
\eea
In particular, the quantities $G_{1,2,3}^+$ and $G_{1,2,3}^-$ are for the Case 1 and Case 2 solutions respectively. The origin of this discontinuity is the same as the $g=0$ case, discussed earlier.

Note that for the above solutions we must have $1-a^2 g^2 > 0$.  When $a^2 g^2 -1>0$, there is another values of $r_+$ where the Case 1 and Case 2 solutions become the same, namely
\be
r_+=\frac{1}{g \sqrt{2(a^2 g^2-1)}}\,.
\ee
The solution is a time machine with $W(r_+)<0$.  In this case, the Gibbs free energies are all continuous crossing one branch of the solutions to the other. There is no asymptotically flat limit of this time machine.

As was discussed earlier, when $g=0$, the Case 1 solution give the supersymmetric BMPV black hole. When $g\ne 0$, the supersymmetric rotating black hole of Gutowski-Reall \cite{Gutowski:2004ez} arises instead from the Case 2 solutions when
\be
a=\fft{g r_+^2}{2 + g^2 r_+^2}\,.
\ee
(This Gutowski-Reall black hole is illustrated in Fig.~\ref{gpsipsi}.) Setting $g=0$ will not give the BMPV black hole, but the extremal static RN black hole with $M=\sqrt3 Q$.

\subsection{Illustration of DBRs from extremal black holes}

\begin{figure}[htp]
\begin{center}
\includegraphics[width=220pt]{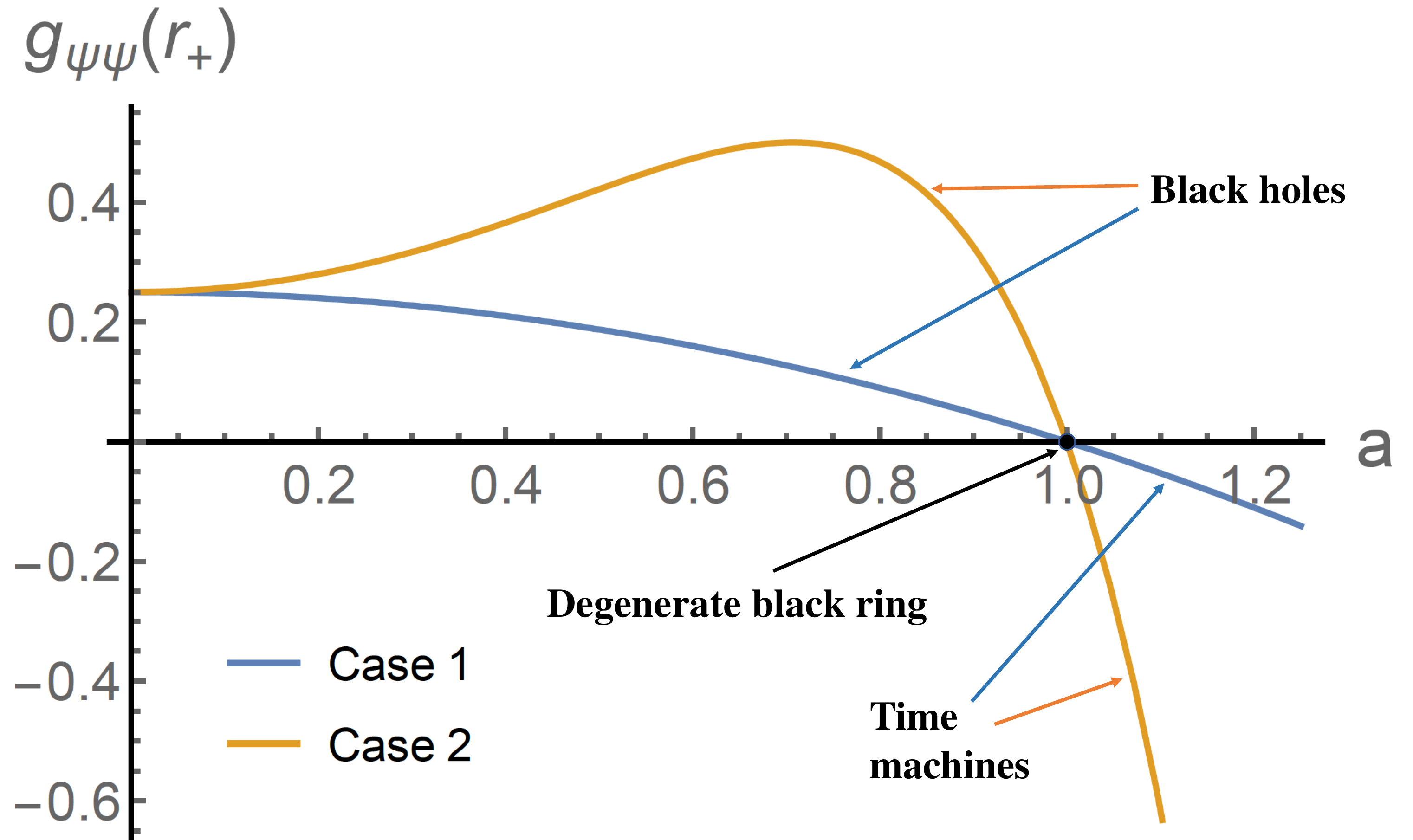}
\includegraphics[width=220pt]{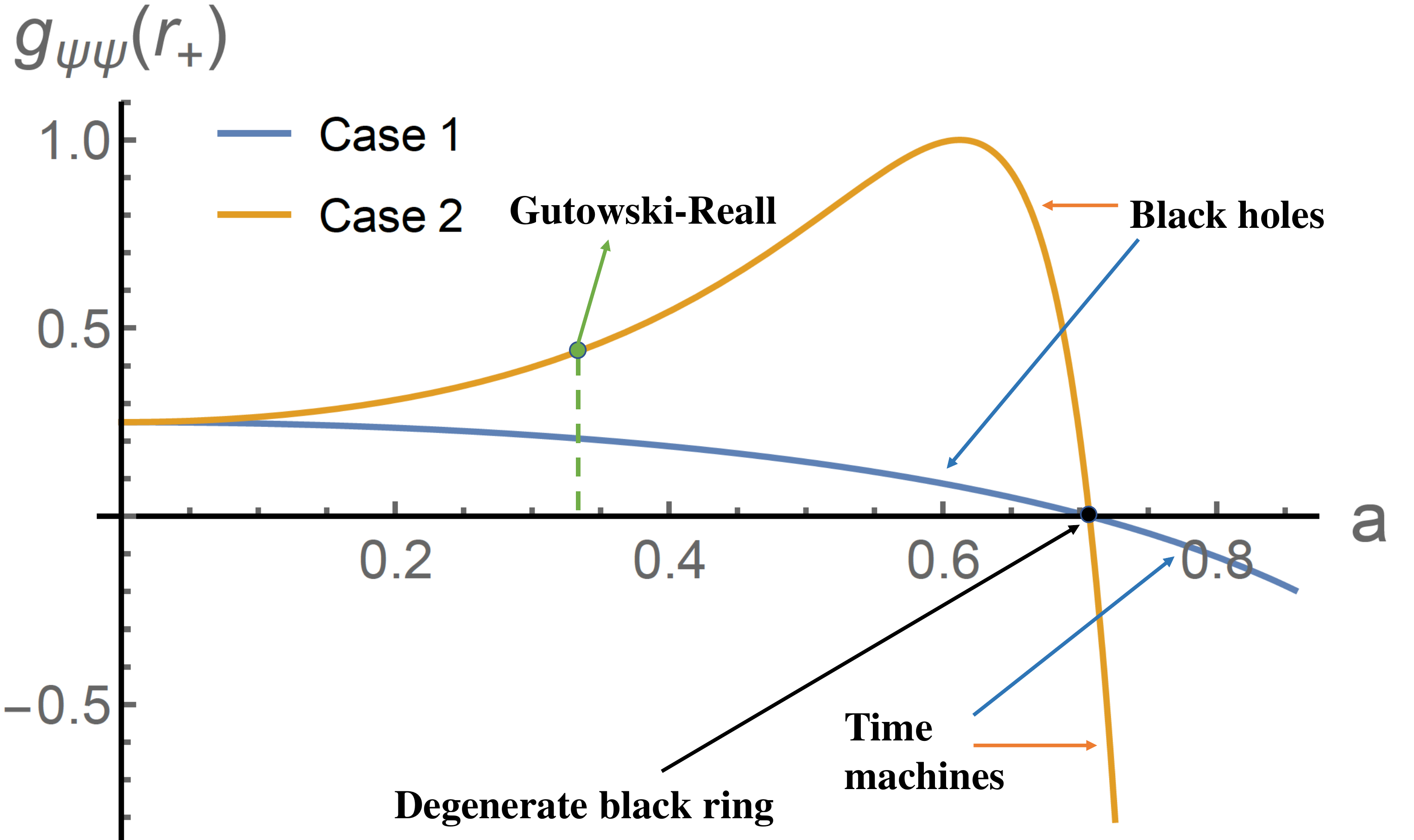}
\end{center}
\caption{\small\it The two branches of the rotating black holes start from the extremal RN black holes at $a=0$ of the same mass and opposite charges and then become the same DBR with $g_{\psi\psi}=0$. They will continue to be time machines if we further increase the parameter $a$.  The global discontinuity of Gibbs free energies arises from the sharp angle at the joining.  We have $r_+=1$ for both graphs, with the left having $g=0$ and the right having $g=1$. The Gutowski-Reall supersymmetric solution sits at $a=1/3$.}
\label{gpsipsi}
\end{figure}

It is instructive to examine the metric component $g_{\psi\psi}(r_+)$ of the two extremal rotating black holes, since the sign of this quantity indicates whether we have a black hole, time machine or DBR. As is illustrated in Fig.~\ref{gpsipsi}, both branches of rotating black holes reduce to the RN black hole when $a=0$. To be precise, the Case 1 solution reduces to the one with $M=-\sqrt3 Q$, whilst the Case 2 solution reduces to the one with $M=+\sqrt3 Q$.  In other words, they have the same mass, but opposite charges. As we increase the value of $a$, the two branches meet at some critical value of $a$, where $g_{\psi\psi}$ vanishes and the DBR emerges. The discontinuity of the Gibbs free energies may arise from the sharp angle at the joining.

\begin{figure}[htp]
\begin{center}
\includegraphics[width=220pt]{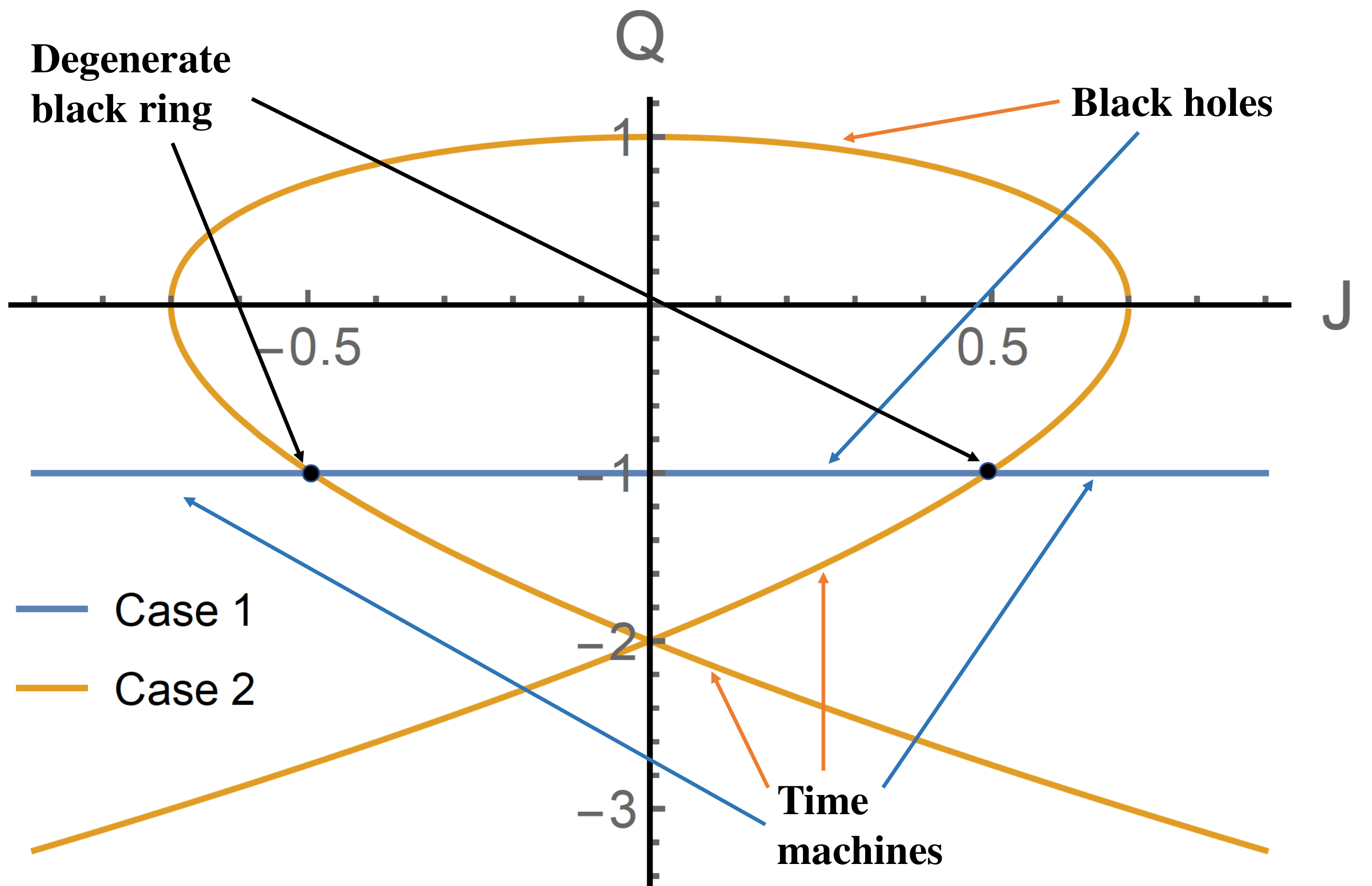}
\includegraphics[width=220pt]{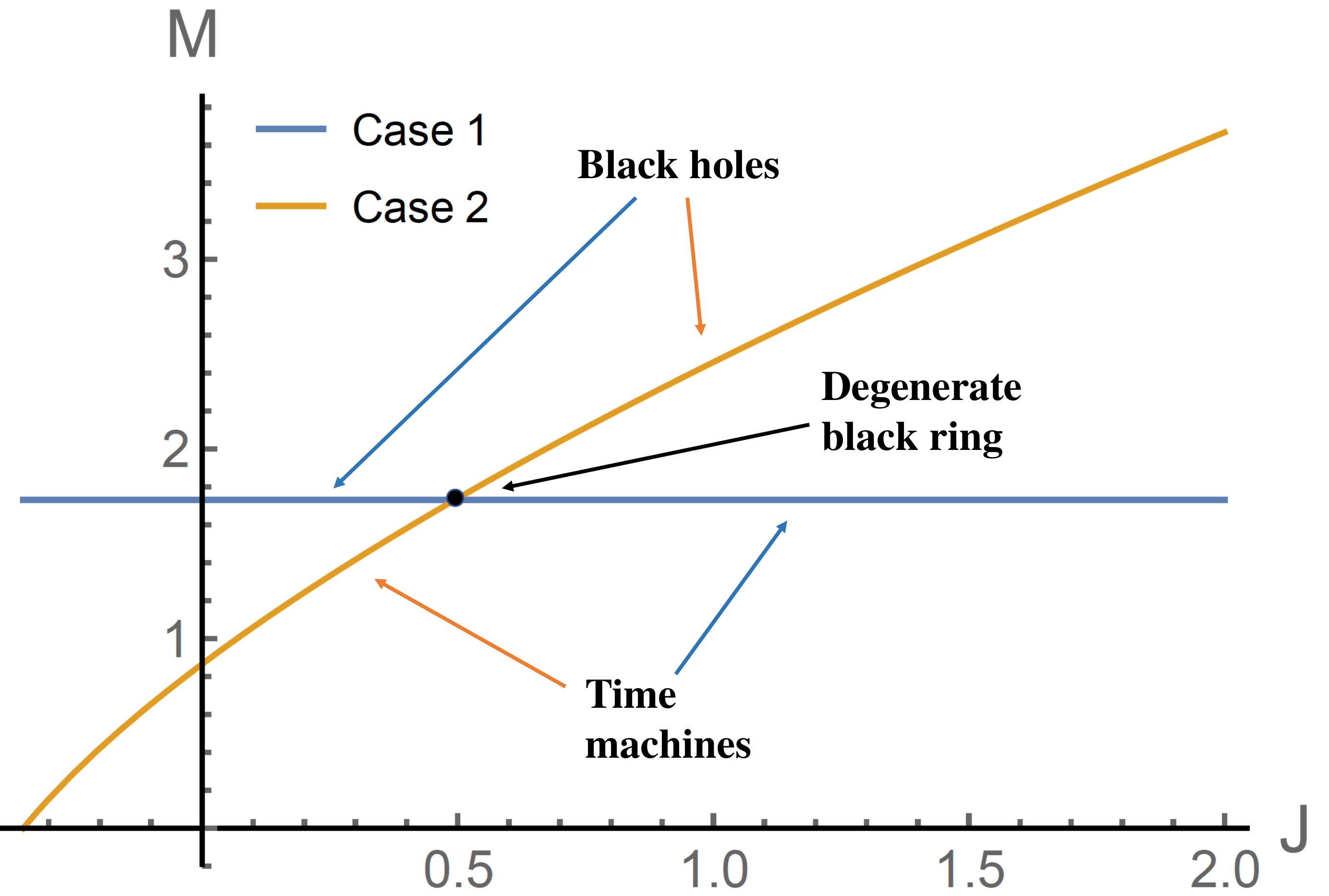}
\end{center}
\caption{\small\it Both plots have $g=0$.  In the left plot, we fix $M=\sqrt3$ and let $J$ take both negative and positive values. In this case, the two branches of extremal rotating black holes join as DBRs at both ends. In the right plot, we fix $Q=-1$. Case 2 gives a black hole when it is over rotating, completely opposite to Case 1.}
\label{mqjplots}
\end{figure}

It is also worth examining how the conserved quantities $(M,Q,J)$ change in the black hole to DBR limit. For simplicity, we consider $g=0$.  We can either fix $M$ or fix $Q$ for both branches of the extremal rotating black holes.  In the left plot of Fig.~\ref{mqjplots}, we fix $M=\sqrt3$, in which case, the extremal RN black holes can have $Q=\pm 1$. We see that the Case 1 solution passes the $Q=-1$ RN black hole, whilst the Case 2 solution passes the $Q=+1$ solution. This splitting of the RN degeneracy by the angular momentum was also observed in \cite{Wu:2009cn}. The two DBR solutions sit at $J=\pm J_{\rm DBR}$ with $J_{\rm DBR}=\frac{2}{3^{3/4} \sqrt{\pi }}\sim 0.495$.  For Case 1, when the solution is under rotating, namely $|J|< J_{\rm DBR}$, it describes a supersymmetric black hole.  When it is over rotating with $|J|> J_{\rm DBR}$, it describes a time machine repulson. The nonsupersymmetric Case 2 solution has much richer structure. Both black holes and time machines can be either under rotating or over rotating.  In particular, there is a rotating time machine with $Q=-2$ without angular momentum.

In the right plot of Fig.~\ref{mqjplots}, we fix $Q=-1$, and draw the mass as a function of $J$. In this case the DBR again corresponds to $J=J_{\rm DBR}$.  What is unusual is that the Case 2 solution describes a black hole when it is over rotating and a time machine when it is under rotating, completely opposite to the Case 1 solution.

\section{DBR from the soliton limit}
\label{sec:soliton}

We can also approach the DBR condition (\ref{dbrcond}) from a different angle, by first requiring $f(r_+)=W(r_+)=0$. This corresponds to the third possibility in \eqref{first3}, which was mentioned in \cite{Cvetic:2005zi} without further elaboration. That the local solution \eqref{localsol} in this case describes regular solitons when $r_L=r_+$ was elaborated in \cite{Andrews:2019hvq}.  We review these solutions here, focusing on the properties that were not addressed in \cite{Andrews:2019hvq}, namely the relations between the conserved quantities $(M,J,Q)$, and how the DBRs arise from the soliton limit.

\subsection{$\Lambda \ne 0$}

The condition that $r_L=r_+\equiv r_0$ can be achieved by requiring
\be
\mu=\frac{r_0^4 \left(a^2+r_0^2\right)}{2 a^4}\,,\qquad
q=-\frac{r_0^4}{a^2}\,.\label{muqsoliton}
\ee
We first consider the case with nonvanishing $\Lambda = -6g^2$, and the spacetime is asymptotic to global AdS. The metric functions are
\bea
W &=& \Big(1 - \fft{r_0^2}{r^2}\Big) \Big(1+\frac{r_0^2}{r^2}+\frac{r_0^6}{a^2 r^4}\Big)\,,
\qquad \omega=-\frac{2 r_0^6}{a \left(a^2 r^2 \left(r^2+r_0^2\right)+r_0^6\right)}\,,\nn\\
f &=& r^2 W\left( \frac{r_0^2 \left(a^2+r^2\right)+r^2 \left(a^2+r^2\right)-r_0^4}{a^2 r^4+a^2 r^2 r_0^2+r_0^6}-\frac{1}{a^2}+g^2\right).
\eea
It is important to note that the quantity in the parenthesis of $f$ should be positive for $r\in [r_0,\infty)$, which requires that
\be
\eta\equiv 1 - (1-a^2 g^2) \fft{r_0^2}{a^2}>0\,.
\ee
Apparently, this condition is always satisfied for $a^2 g^2\ge 1$; otherwise, it further requires a small enough $r_0$.

The mass, angular momentum and electric charge are not independent, with $Q$ precisely given by \eqref{Q(J)}. The mass is
\be
M = \frac{J \left(a^2 g^2+3\right)}{2 a}+\frac{\sqrt[3]{\pi } J^{2/3} \left(3-a^2 g^2\right)}{2\ 2^{2/3}}\,.
\ee
In other words, the charge $Q$ is negative and depends on $J$, but the mass $M$ appears to be an independent parameter, because it depends on the free parameter $a$.

However, there is an additional constraint as one studies the geometry at the coordinate singularity $r=r_0$.  To see this, we let $r=r_0 + \ft14 \rho^2$ and in the vicinity of $\rho=0$, the metric becomes
\bea
ds^2 &=&-\eta dt^2 + \frac{a^2 r_0}{2 \eta \left(2 a^2+r_0^2\right)} \Big(d\rho^2 + \ft14\kappa_{\rm E}^2\, \rho^2 \sigma_3^2\Big) + \ft14 r_0^2 d\Omega_2^2\,,
\eea
where the ``Euclidean surface gravity'' is
\be
\kappa_{\rm E} = \frac{(2 a^2+r_0^2)\sqrt{\eta}}{a^2}=\frac{\left(2 \sqrt[3]{\pi } a+2^{2/3} \sqrt[3]{J}\right) \sqrt{2^{2/3} \sqrt[3]{J} \left(a^2 g^2-1\right)+\sqrt[3]{\pi } a}}{\sqrt{\pi } a^{3/2}}\,.
\ee
Thus in order to avoid the conic singularity at $r=r_0$, we must have $\Delta \psi = 4\pi/\kappa$.  On the other hand, as was discussed in section \ref{sec:local}, the period of $\psi$ is constrained by the 3-sphere or lens space topology as in \eqref{psiperiod}, we thus must have
\be
\kappa_{\rm E}=k\,,\qquad k=1,2,3,\cdots\,.\label{kappacons}
\ee
Consequently, the integration constant $a$, and hence the mass is not a free parameter, but should be determined by the angular momentum $J$ and the discrete topological parameter $k$. Specifically, $a$ is determined by the fourth order polynomial equation
\be
4 (2 \pi )^{2/3} a^4 g^2 \sqrt[3]{J}-a^3 \left(\pi  \left(k^2-4\right)-8 \sqrt[3]{2 \pi } g^2 J^{2/3}\right)+4 a^2 g^2 J- 6 \sqrt[3]{2 \pi }\, a J^{2/3}-4J=0\,.
\ee
For large $J$, the function $a(J)$ approaches $1/g$, given by
\be
a=\frac{1}{g}-\frac{\sqrt[3]{\pi } \sqrt[3]{\frac{1}{J}}}{2 \left(2^{2/3} g^2\right)}+\frac{\pi ^{2/3} \left(\frac{1}{J}\right)^{2/3}}{16 \sqrt[3]{2} g^3}+\frac{\pi  k^2}{8 g^4 J}+O\left(J^{-\fft43}\right)\,.
\ee
The behavior of $a(J)$ for small $J$ depends on the values of $k$:
\bea
k=1:&& a= \frac{2^{2/3} c \sqrt[3]{J}}{\sqrt[3]{3 \pi }}\nn\\
&&\qquad -\frac{4 g^2 J \left(3 \sqrt[3]{3} c^2+4 c \left(3^{2/3}+6 \sqrt[6]{3} \cos \left(\frac{\pi }{18}\right)\right)+12+24 \sqrt{3} \cos \left(\frac{\pi }{18}\right)\right)}{27 \pi  \left(\sqrt[3]{3} c^2-1\right)} + \cdots\nn\\
&&\phantom{a}= 1.23251 \sqrt[3]{J}-2.04349 g^2 J+\cdots\,,\qquad \left(c=\sqrt[3]{1+2 \sqrt{3} \cos \left(\frac{\pi }{18}\right)}\right)\,;\nn\\
k=2:&& a= \frac{\sqrt[3]{3} \sqrt[9]{J}}{2^{4/9} \sqrt[9]{\pi } g^{2/3}} - \frac{2\ 2^{2/3} \sqrt[3]{J}}{9 \sqrt[3]{\pi }} + \frac{13 g^{2/3} J^{5/9}}{81\ 2^{2/9} \sqrt[3]{3} \pi ^{5/9}}+\cdots\,;\nn\\
k\ge 3:&& a=\frac{\sqrt[3]{\pi } \left(k^2-4\right)}{4\ 2^{2/3} g^2 \sqrt[3]{J}} -\frac{2^{2/3} \sqrt[3]{J}}{\sqrt[3]{\pi }} - \frac{4 g^2 J \left(k^2-16\right)}{\pi  \left(k^2-4\right)^2} + \cdots\,.
\eea
For $k=1,2$, $a(J)$ is a monotonic function increasing from 0 to $1/g$ as $J$ runs from zero to infinity.  When $k\ge 3$, $a(J)$ has a positive minimum with $a(0)\rightarrow \infty$ and $a(\infty)=g^{-1}$.

The mass $M(J)$ is a monotonic function of the angular momentum for all $k$, with small $J$ expansion:
\be
M=
\left\{
  \begin{array}{ll}
    2.60098 J^{2/3}+1.93329 g^2 J^{4/3}+O\left(J^2\right), &\quad k=1; \\
    \frac{3 \sqrt[3]{\pi } J^{2/3}}{2\ 2^{2/3}}+\frac{3\ 3^{2/3} \sqrt[9]{\pi } g^{2/3} J^{8/9}}{4\ 2^{5/9}}+\frac{7 g^{4/3} J^{10/9}}{2\ 2^{4/9} 3^{2/3} \sqrt[9]{\pi }}+O\left(J^{4/3}\right), &\quad k=2; \\
   -\frac{\pi  \left(k^2-4\right)^2}{128 g^2}+\frac{3 \sqrt[3]{\pi } J^{2/3} k^2}{8\ 2^{2/3}}-\frac{3 J^{4/3} \left(g^2 \left(k^2-8\right)\right)}{2 \left(\sqrt[3]{2 \pi } \left(k^2-4\right)\right)}+O\left(J^2\right), &\quad k\ge 3\,.
  \end{array}
\right.
\ee
Note that mass becomes negative for $k\ge 3$ for small $J$.  In fact the $J=0$ solution becomes the static AdS solitons of negative mass from Einstein metrics studied in \cite{Clarkson:2005qx}.  For large $J$, the $M(J)$ function takes a universal form:
\be
M=2 g J+\frac{3 \sqrt[3]{\pi } J^{2/3}}{2\ 2^{2/3}}+\frac{3 \pi ^{2/3} \sqrt[3]{J}}{8 \sqrt[3]{2} g}-\frac{(2k^2+1)\pi }{16 g^2}+O\left(J^{-\fft13}\right)\,.\label{MJsolitonk}
\ee
Note that for general $k$, the mass and charge should be replaced by $M_k$ and $J_k$ defined in \eqref{mjqk}.

Intriguingly, if we take $k=0$, so that the Euclidean surface gravity $\kappa_{\rm E}=0$, we have $\eta=0$, and the local solution reduces precisely to the AdS DBR studied in section \ref{sec:dring}.  Thus the DBR solution can also be viewed as the limiting case of the solitons, although the parameter $k$ is a discrete variable. However, taking $k=0$ does not mean that $\psi$ becomes a real line.  The structure of the Killing horizon of the soliton now becomes the event horizon in this limit and there is no longer a further restriction on the period $\psi$, which can be $\Delta \psi=4\pi/k$ for any $k$.  We prefer to choose $k=1$ so that the solution is asymptotic to the global AdS$_5$.  In other words, the $k=0$ limit of the soliton is singular such that the topology changes completely and consequently the asymptotic AdS spacetime remains global, the same as the $k=1$ soliton.

\begin{figure}[htp]
\begin{center}
\includegraphics[width=270pt]{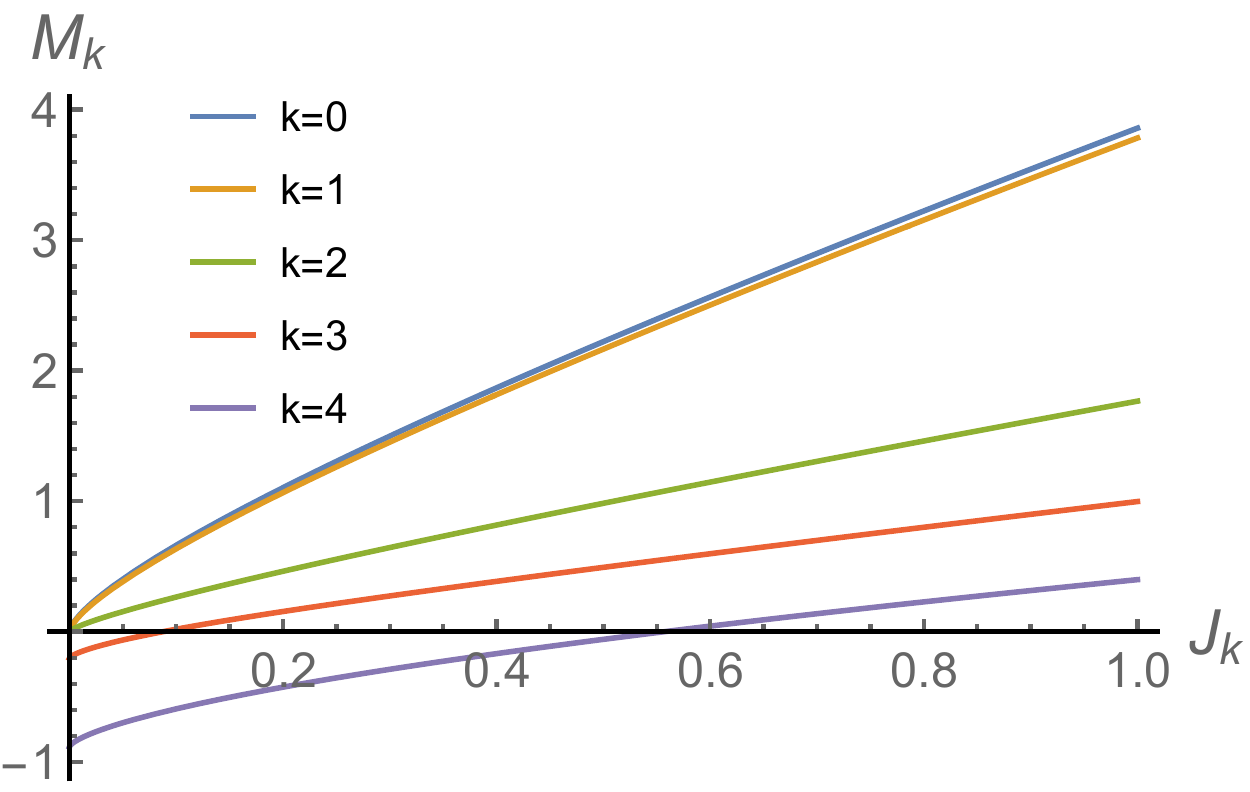}
\end{center}
\caption{\small\it Here are the mass-angular momentum relations of the AdS solitons ($g=1$) with $\mathbb{R}\times S^3/\mathbb{Z}_k$ boundaries. The $k=0$ case corresponds to the degenerate AdS black ring, whose asymptotic infinity is the global AdS, the same as the $k=1$ soliton. The soliton mass can be negative for $k\ge 3$.}
\label{MJplots}
\end{figure}

Note that the large $J$ expression of the mass in \eqref{mjadsring} is precisely the special $(k=0)$ case of the more general expression \eqref{MJsolitonk}. In Fig.~\ref{MJplots}, we give the mass-angular momentum relation $M_k(J_k)$ for $k=0,1,2,3,4$, where $(M_k,J_k)$ are defined by \eqref{mjqk}, with the $k=0$ case corresponding to the AdS DBR.  Since all these solutions have the same charge $Q(J)$, it is thus sensible to compare their mass for given $J$. However, the comparison makes sense only for solutions with the same asymptotic infinity.
Both the degenerate AdS black ring and the $k=1$ soliton hav the same global AdS as their asymptotic infinity, and we find that the soliton has lower mass than the corresponding black ring. Since the period $\Delta \psi=4\pi/k$ for the black ring is allowed for any integer $k$, we can compare the mass of degenerate AdS black ring with $S^3/\mathbb{Z}_k$ and we find that it is always higher than the mass of the corresponding $k$-soliton.

\subsection{$\Lambda=0$}

We now analyse the soliton solutions when $g=0$, such that the solutions are asymptotic to the flat spacetime. It is straightforward to set $g=0$ for both Case 1 and Case 2 solutions and also the formulae of their conserved quantities $(M,J,Q)$.  The condition of being a soliton \eqref{muqsoliton} is independent of $g$. The subtlety lies on the regularity condition $\kappa_{\rm E}=k$. When $g=0$, we have
\be
\kappa^2_{\rm E} - k^2 = 4-k^2  -\frac{r_0^6}{a^6}-\frac{3 r_0^4}{a^4}=0\,.
\ee
This condition cannot be satisfied unless $k=1$ or 2.  When $k=2$, we must have $a=\infty$, which implies that $Q=0=J$.  The solution becomes simply a direct product of time and the Eguchi-Hanson instanton. The more interesting case is $k=1$.  We have
\be
\fft{r_0^2}{a^2}=2 \cos \left(\frac{\pi }{9}\right)-1\,.
\ee
It follows that we obtain the asymptotically Minkowskian rotating soliton with \eqref{Q(J)} and
\be
M=\frac{3 \sqrt[3]{\pi } \cos \left(\frac{\pi }{9}\right)}{2^{2/3}} J^{\fft23}\,.\label{mjqsoliton2}
\ee
Thus we see that $M\sim -Q$, but not the supersymmetric condition $M=- \sqrt3 Q$, which gives the BMPV solution. Furthermore, as we have remarked in section \ref{sec:dring}, the mass of the soliton is smaller than the corresponding DBR with the same $Q(J)$.

\section{Conclusions}
\label{sec:conclusion}

In this paper, we analysed the charged rotating solutions \cite{clpd5sol1} in five-dimensional minimal supergravity, with or without the cosmological constant. The general solution has three nontrivial integration constants, parameterizing the mass $M$, electric charge $Q$ and two equal angular momenta $J$. For suitable $(M,Q,J)$, we found that new spacetime geometries could arise that could be best described degenerate black rings, or DBRs. The metric functions of the DBRs satisfy the condition \eqref{dbrcond} and thus they have one free parameter $J$; mass and charge are functions of $J$.  The relation $Q(J)$ \eqref{Q(J)} is universal and does not depend on the cosmological constant, but $M(J)$ does. The DBRs are extremal black objects that are asymptotic to either Minkowski or global AdS spacetimes, and the solutions can avoid the curvature power-law singularity by introducing an orbifold singularity on its AdS$_3\times S^2$ near-horizon geometry. Two factors help to evade the no-go theorems of \cite{Grover:2013hja,Khuri:2017zqg}: our solutions are not supersymmetric; the ring sizes are all degenerate.

We studied two routes that lead to the DBR solutions from the general class of solutions.  One is to consider extremal rotating black holes that satisfy $f(r_+)=f'(r_+)=0$. It turns out that there are two branches of such solutions and they join at the point of parameter space where the DBR emerges.  At the DBR limit from the two generally different extremal black holes, the mass or the Helmholtz free energy is continuous, but the Gibbs free energies are not. We showed that the source of this discontinuity was the $FFA$ term, which was not gauge invariant, even though its contribution to the equations of motion was. The physical implication of this unusual global discontinuity deserves further investigation.

We can also take the DBR limit from the rotating solitons that satisfy $f(r_+)=W(r_+)=0$.  These solitons are characterized by the condition that the Euclidean surface gravity $\kappa_{\rm E}$ is equal to the topological parameter $k$, where the level surfaces of the spatial slice is $S^3/\mathbb{Z}_k$.  The DBR emerges locally as the $k=0$ limit; however, owing to the change of the topology, the DBR has the same asymptotic global AdS$_5$ or Mink$_5$ as the $k=1$ soliton.  The DBRs and solitons have the same $Q(J)$ relation and the solitons have the smaller mass for given charge.

Five-dimensional gauged supergravity has an origin of D3-brane in type IIB string theory and its AdS/CFT dual to four-dimensional super Yang-Mills are expected to be exact. Thus our DBRs may have implications in the AdS/CFT correspondence.  The DBRs and AdS solitons are dual to spinning operators whose conformal weight, $U(1)$ global charge and spin are proportional to $(M,Q,J)$ of these bulk solutions.  These operators are restricted by the negative $Q(J)$ relation \eqref{Q(J)}, and their mass for large $J$ takes the form
\be
M=c_{\fft33} J + c_{\fft23} J^{\fft23} + c_{\fft13} J^{\fft13} + c_0 + \cdots\,,
\ee
with $c_{\fft33}=2g$ being universal. (Of course, we can also give the alternative but equivalent expression as $M(Q)$.) The bulk theory gives the explicit coefficients $c$'s, with the $k=1$ soliton being the ground state. It is of great interest whether this conclusion can be established in the dual CFT.

\section*{Acknowledgement}

We are grateful to Jiaju Zhang for the naming of the degenerate black ring. The paper is supported in part by the National Natural Science Foundation of China (NSFC) grants No. 11875200 and No. 11935009.

\section*{Appendix}
\appendix

\section{Black holes and time machines}
\label{app}

In this appendix, we give some global analysis of the local solution \eqref{localsol}. Here we focus on the first two cases of \eqref{first3}.  In particular, we establish the first law of black hole thermodynamics and corresponding thermodynamical variables that play an important role in the discussions of the main text.

\subsection{Charged rotating AdS black holes}

For appropriate parameters, the metric function $f$ has real roots and we let $r_+>0$ be the largest one. The condition $f(r_+)=0$ implies
\be
\mu=\fft{1}{2a^2\tilde \eta}\Big(2 a^2 q+q^2+r_+^4 +g^2 \left(r_+^6-a^2 q \left(q-2 r_+^2\right)\right) \Big)\,,\qquad \tilde \eta =\frac{r_+^2 \left(1-a^2 g^2\right)}{a^2}-1\,.
\ee
It follows from the fact
\be
W(r_+) = \frac{\left(a^2 q+r_+^4\right){}^2}{\tilde \eta  a^2 r_+^6}
\ee
that if $\tilde \eta>0$, $r=r_+$ is the event horizon, and the VLS and hence CTCs are inside.  The null Killing vector on the horizon is given by
\be
\ell = \fft{\partial}{\partial t} + 2 \Omega_+ \fft{\partial}{\partial \psi}\,,\qquad
\Omega_+= \frac{a \left(g^2 r_+^4+q+r_+^2\right)}{r_+^4+a^2 q}\,.
\ee
Note that the angular velocity $\Omega_+$ is calculated from the metric that is non-rotating asymptotically. The electric potential is given by
\be
\Phi = \ell^\mu A_\mu\Big|_{r_+} = \fft{\sqrt3 a^2 q \tilde \eta}{r_+^4 + a^2 q}\,.\label{phipot}
\ee
This result arises from the gauge choice that the quantity $\ell^\mu A_\mu$ vanishes at the asymptotic infinity. As was explained in section \ref{sec:phase}, the action is not gauge invariant owing to the $FFA$ term and it is necessary to require that $A^2$ is non-divergent on and outside of the horizon. This can be achieved by shifting the temporal component with some appropriate constant such that $\ell^\mu A_\mu$ vanishes on the horizon.
It is then straightforward verify that the Euclidean action will yield the correct Gibbs free energy.

The surface gravity $\kappa$ is given by $\kappa^2 = - g^{\mu\nu} \partial_\mu \ell^2
\partial_\nu \ell^2/(4\ell^2)$ at $r=r_+$ and the temperature is
\be
T=\fft{2\pi}{\kappa} = \fft{\left(q+r_+^2\right) \left(r_+^2-2 a^2-q\right)+2 g^2 \left(a^2 q^2-2 a^2 r_+^4+r_+^6\right) -a^2 g^4 \left(a^2 q^2+2 r_+^6\right) }{2 \pi  \left(a^2 q+r_+^4\right)\sqrt{\tilde \eta}}\,.\label{apptemp}
\ee
The entropy is the standard one quarter of the area of the horizon
\be
S=\frac{\pi ^2 \left(r_+^4 +a^2 q\right)}{2 a \sqrt{ \tilde \eta }}\,.
\ee
It is then straightforward to obtain the first law of black hole thermodynamics and the
corresponding Smarr relation
\be
dM=TdS + \Phi dQ -2 \Omega_+ dJ + V_{\rm th} dP_{\rm th}\,,\qquad
M=\ft32 TS + \Phi Q + 3 \Omega_+ J - V_{\rm th} P_{\rm th}\,.
\ee
Note that here for completeness, we introduced the black hole thermodynamical pressure and the corresponding volume \cite{Kastor:2009wy,Cvetic:2010jb}
\be
P_{\rm th} =\fft{3g^2}{4\pi}\,,\qquad V_{\rm th} =
\fft{\pi^2\Big(3 r_+^6+a^2 \left(2 q^2+4 q r_+^2-r_+^4\right)-a^2 g^2 \left(2 a^2 q^2+r_+^6\right)\Big)}{6a^2 \tilde \eta}\,.
\ee
They do not play any role in our discussions in the main text.

The black hole becomes extremal when $T=0$, in which case, the mass $M$ becomes a function of $J$ and $Q$.  (As was elaborated in section \ref{sec:phase}, two branches of such solutions exist, starting as the RN black holes of the same mass but opposite charges, and joining as the DBR at the other end.) An additional constraint is needed for the solution to become supersymmetric \cite{Gutowski:2004ez}, namely
\be
M-2g J - \sqrt3 Q=0\,.
\ee
This implies that the solution are described by only one integration constant, with
\be
\mu = \frac{1}{2} r_+^2 \left(g^4 r_+^4+3 g^2 r_+^2+2\right)\,,\qquad q=\frac{1}{2} g^2 r_+^4+r_+^2\,,\qquad a=\frac{g r_+^2}{g^2 r_+^2+2}\,.
\ee
The mass, angular momentum and electric charge are now
\be
M=\frac{1}{8} \pi  r_+^2 \left(4 g^4 r_+^4+9 g^2 r_+^2+6\right)\,,\quad J=\frac{1}{8} \pi  g r_+^4 \left(2 g^2 r_+^2+3\right)\,,\quad Q=\frac{1}{8} \sqrt{3} \pi  r_+^2 \left(g^2 r_+^2+2\right)\,.
\ee
For large $J$, we have
\bea
M -2g J = \sqrt3 Q=\frac{3 \sqrt[3]{\pi } J^{2/3}}{2\ 2^{2/3}}  +\frac{3 \pi ^{2/3} \sqrt[3]{J}}{4 \sqrt[3]{2} g} -\frac{3 \pi }{32 g^2} +O\left(J^{-\fft13}\right)\,.
\eea
This mass is larger than both the DBR and the $k=1$ soliton, but the comparison may not be valid since they have different $Q(J)$.

\subsection{Time machines}

When $\tilde \eta$ is negative, then $W(r_+)$ is negative, implying that the VLS is outside the Killing horizon, and the solution describes a time machine \cite{Cvetic:2005zi}. In this case, the ``temperature'' \eqref{apptemp} becomes purely imaginary and regularity on the Killing horizon requires that the real time have to be in general periodic.  The geodesics is then complete on and outside the Killing horizon and the curvature power-law singularity at $r=0$ cannot be reached.  Intriguingly, rotating black holes, even such as the Kerr metrics, become regular time machines when the mass is negative \cite{Feng:2016dbw}.

Charged AdS time machines can be supersymmetric, and the supersymmetric condition can be satisfied with $q=-\mu$, for which one has \cite{Klemm:2000vn}
\be
M=-\sqrt3 Q=\ft34\pi \mu\,,\qquad J=\ft14\pi \mu a\,.
\ee
We can impose further that $f$ has a double root $r_0$ so that there would be no periodic condition on the real time coordinate $t$. This can further constrain the parameter $a$, so that the solution has only one integration constant left:
\be
\mu =-q=r_0^2 + \ft32g^2 r_0^2\,,\qquad a= \frac{r_0 \sqrt{9 g^2 r_0^2+4}}{3 g^2 r_0^2+2}\,,
\ee
The mass and the electric charge are now both functions of $J$:
\be
M=-\sqrt3 Q = \frac{3}{8} \pi  r_0^2 \left(3 g^2 r_0^2+2\right)\,,\qquad
J=\frac{1}{8} \pi  r_0^3 \sqrt{9 g^2 r_0^2+4}\,.
\ee
We thus have
\be
M=-\sqrt3Q = 3g J + \sqrt{\fft{2\pi J}{3g}}- \fft{\pi}{36 g^2} + O\left(J^{-\fft12}\right)\,.
\ee
This relation shows that it is a very different class of solutions from the ones we discussed in the main text.  Specifically, the solution is given by
\bea
f &=& \Big(1 - \fft{r_0^2}{r^2}\Big)^2 \big(1 + g^2 (r^2 + 2r_0^2)\big)\,,\qquad
W=1-\frac{r_0^6 \left(9 g^2 r_0^2+4\right)}{4 r^6}\,,\nn\\
\omega &=& -\frac{2 r_0^3 \sqrt{9 g^2 r_0^2+4} \left(2 r^2-2 r_0^2-3 g^2 r_0^4\right)}{4 r^6-4 r_0^6-9 g^2 r_0^8}\,,\qquad A=\frac{3 r_0^6 \left(9 g^2 r_0^2+4\right)}{16 r^4}\tilde\sigma_3\,.
\eea
The VLS $r_L$ where $W(r_L)=0$ is outside the Killing horizon $r_0$, with
\be
\fft{r_L}{r_0} = \sqrt[6]{1+\ft{9}{4} g^2 r_0^2}>1\,.
\ee
When $g=0$, it reduces to the DBR that is asymptotic to Minkowski spacetime, analysed in section \ref{sec:dring}.

\end{document}